\documentclass[aps,prl,reprint,amsmath,amssymb,showpacs]{revtex4-1}
\usepackage[ansinew]{inputenc}
\usepackage{graphicx}
\usepackage{textcomp}
\usepackage{pifont}
\usepackage{enumerate}
\usepackage{hyperref}
\usepackage[all]{hypcap}
\input{glyphtounicode}
\newcommand{\bra}[1]{\langle #1 |}
\newcommand{\ket}[1]{|#1\rangle}

\begin{document}

\title{Heralded Storage of a Photonic Quantum Bit in a Single Atom}
\author{Norbert Kalb}
\altaffiliation[Present address: ]{Kavli Institute of Nanoscience Delft, Delft University of Technology, PO Box 5046, 2600 GA Delft, The Netherlands}
\author{Andreas Reiserer}
\altaffiliation[Present address: ]{Kavli Institute of Nanoscience Delft, Delft University of Technology, PO Box 5046, 2600 GA Delft, The Netherlands}
\author{Stephan Ritter}
\email{stephan.ritter@mpq.mpg.de}
\author{Gerhard Rempe}
\affiliation{Max-Planck-Institut f\"ur Quantenoptik, Hans-Kopfermann-Strasse 1, 85748 Garching, Germany}

\begin{abstract}
Combining techniques of cavity quantum electrodynamics, quantum measurement, and quantum feedback, we have realized the heralded transfer of a polarization qubit from a photon onto a single atom with 39\,\% efficiency and 86\,\% fidelity. The reverse process, namely, qubit transfer from the atom onto a given photon, is demonstrated with 88\,\% fidelity and an estimated efficiency of up to 69\,\%. In contrast to previous work based on two-photon interference, our scheme is robust against photon arrival-time jitter and achieves much higher efficiencies. Thus, it constitutes a key step toward the implementation of a long-distance quantum network.
\end{abstract}

\pacs{03.67.Hk, 42.50.Ex, 42.50.Pq}

\maketitle

Optical photons are ideal information carriers for quantum networks on a global scale \cite{kimble_quantum_2008, briegel_quantum_1998, duan_long-distance_2001, cirac_quantum_1997, ritter_elementary_2012}. In the envisioned quantum network, quantum information will be reversibly transferred between nodes via the controlled emission, propagation, and absorption of an optical photon. Unfortunately, all three of these processes suffer from losses and inefficiencies, making the information transfer probabilistic and hindering the implementation of large quantum networks. The resulting randomness can be overcome with a herald which, by means of a suitable measurement, unambiguously signals the successful transfer of information between network nodes.

The first process mentioned above---generation of a photon---can be signalled by detecting one of the two photons emitted in a probabilistic two-photon process \cite{duan_long-distance_2001, eisaman2011}. The second step---the successful transmission of a photon, for example through a long optical fiber---can be heralded using a non-destructive photon detector \cite{reiserer_nondestructive_2013}. However, and in spite of first experiments \cite{tanji_heralded_2009, kurz_high-fidelity_2014}, efficient schemes for the most important final step---the heralded absorption of an incoming photon---are still missing. A way around \cite{duan_long-distance_2001} is to use the techniques of linear optical quantum computing \cite{kok2007}, especially the optical Bell-state measurement (BSM) with a locally produced ancillary photon. This measurement scheme has become a workhorse, for example, to teleport photonic quantum states into a quantum memory \cite{chen_memory-built-quantum_2008, hammerer_quantum_2010, bao_quantum_2012, nolleke_efficient_2013, gao_quantum_2013} or to realize entanglement and quantum-state transfer between remote memories \cite{duan_colloquium:_2010, sangouard_quantum_2011, hammerer_quantum_2010, bao_quantum_2012, nolleke_efficient_2013, hofmann_heralded_2012, bernien_heralded_2013}, but still faces fundamental limitations regarding efficiency and robustness.

To understand this, note that an optical BSM for single-photon qubits using linear optics elements requires the generation of a photon followed by the interference and subsequent detection of two photons. Its success probability is therefore limited \cite{Calsamiglia2001} to $\frac{1}{2}\eta_\mathrm{gen}\eta_\mathrm{det}^2$, where $\eta_\mathrm{gen}$ is the photon generation efficiency and $\eta_\mathrm{det}$ is the quantum efficiency of the employed single-photon detectors. The product of three typically small numbers together with the impossibility to unambiguously identify all of the four Bell states keeps the BSM efficiency small. Another drawback is that an optical BSM requires photons which are indistinguishable in all degrees of freedom except those used for information encoding, e.g. the polarization. Thus, one needs perfect control over both the arrival time of the photons and their spectral properties, which is hard to achieve in many experiments and even harder outside the laboratory.

Here, we overcome all of these limitations and demonstrate the heralded, highly efficient transfer of a photonic polarization qubit onto a single rubidium atom trapped at the center of an optical cavity. Toward this end, we employ an atom-photon quantum gate \cite{reiserer_quantum_2014} that is based on photon reflection from the cavity. The storage efficiency is given by $R\eta_\mathrm{det}$, where $R$ is the reflectivity of the atom-cavity system on resonance. Our experiment exhibits an average $R=(69\pm2)$\,\% and achieves an efficiency of $(39\pm4)\,\%$, which is a factor of $2R/(\eta_\mathrm{gen}\eta_\mathrm{det})$, i.e., more than 4 times, higher than what can be achieved by using an optical BSM with state-of-the-art photon sources ($\eta_\mathrm{gen}=0.6$) \cite{mucke_generation_2013} and the best commercially available single-photon detectors [$\eta_\mathrm{det}=0.56(5)$ at 780\,nm]. Even with the unrealistic assumption of a perfect single-photon source ($\eta_\mathrm{gen}=1$) and perfect single-photon detectors ($\eta_\mathrm{det}=1$), employed both in our realization and in the optical BSM, we improve by a factor of $1.4$, thereby outperforming any optical BSM. In addition, our scheme is inherently more robust against variations in the properties of the transmitted photons and, therefore, ideally suited for the implementation of long-distance protocols under realistic, real-world conditions.

To explain the working principle of our experiment, both the atomic and the photonic quantum state are described as an effective two-level system, with the basis states $\ket{\uparrow_z}$ and $\ket{\downarrow_z}$. The photonic qubit is encoded in the polarization of a weak coherent laser pulse, where left-(right-)circular polarization encodes the state $\ket{\downarrow_z^p}$($\ket{\uparrow_z^p}$). The atomic qubit is encoded in the $\ket{F,m_F}$ states $\ket{\downarrow_z^a} \equiv \ket{1,1}$ and $\ket{\uparrow_z^a} \equiv \ket{2,2}$, where $F$ denotes the hyperfine ground state of $^{87}$Rb atoms and $m_F$ its projection onto the quantization axis.
To simplify the notation, we define the following basis states:
\begin{align*}
\ket{\uparrow_x} & \equiv \frac{1}{\sqrt{2}}\left(\ket{\uparrow_z} + \ket{\downarrow_z}\right), & \ket{\downarrow_x} & \equiv \frac{1}{\sqrt{2}}\left(\ket{\uparrow_z} - \ket{\downarrow_z}\right),\\
\ket{\uparrow_y} & \equiv \frac{1}{\sqrt{2}}\left(\ket{\uparrow_z} + i \ket{\downarrow_z}\right), & \ket{\downarrow_y} & \equiv \frac{1}{\sqrt{2}}\left(\ket{\uparrow_z} - i \ket{\downarrow_z}\right).
\end{align*}

\begin{figure}
\includegraphics[width=\columnwidth]{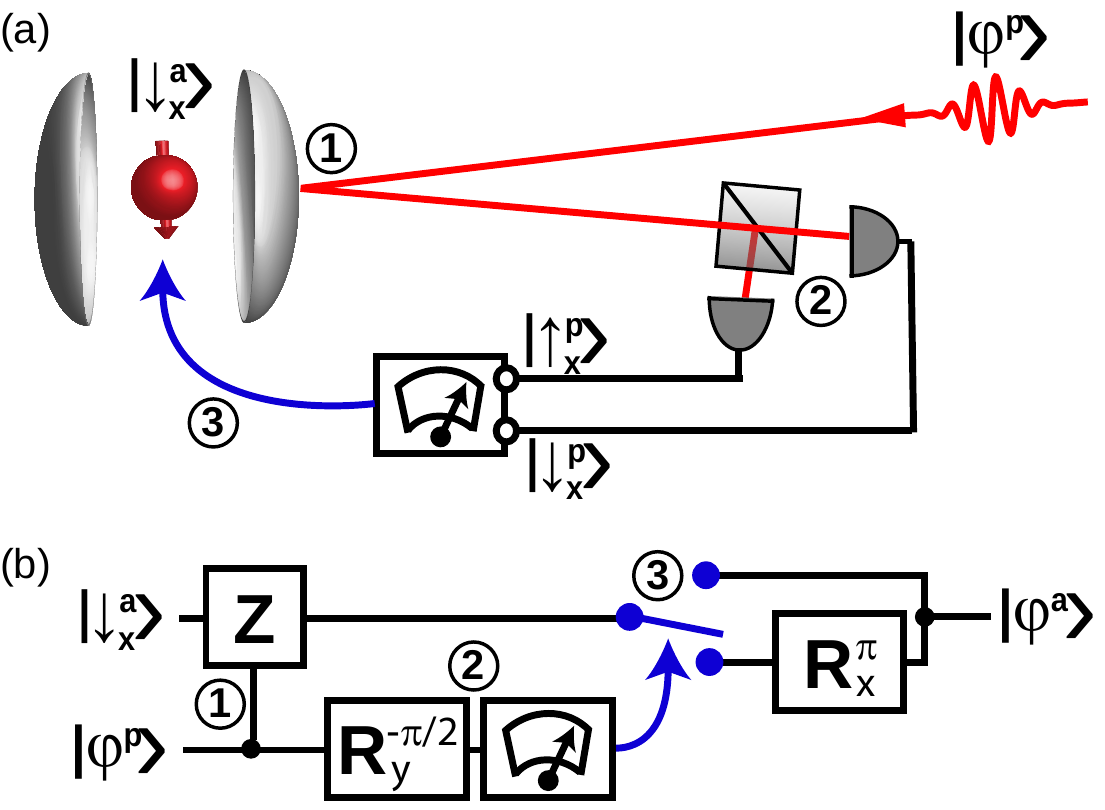}
\caption{\label{fig:setup}
(color online). (a) Heralded storage scheme and (b) corresponding quantum circuit. The atom (red sphere with arrow), trapped inside an optical cavity (gray spherical mirrors), is initialized in $\ket{\downarrow_x^a}$. The photon, whose polarization $\ket{\varphi^p}$ is to be stored, is reflected from the atom-cavity setup \textcircled{1}, thereby performing a controlled-phase quantum gate. The action of this gate (controlled-$Z$) is to introduce a phase shift of $\pi$ to all atom-photon states but $\ket{\uparrow_z^a}\ket{\uparrow_z^p}$. The photon is subsequently detected in the x-basis \textcircled{2} using a polarizing beam splitter (gray cube) and two single-photon counters (gray). This photon detection heralds the state transfer. To complete it, quantum feedback in the form of a state rotation (blue arrow, \textcircled{3}) is applied to the atom, conditioned on the measurement outcome $\ket{\downarrow_x^p}$ or $\ket{\uparrow_x^p}$.}
\end{figure}

The atom-photon interaction is based on reflecting the photon from the cavity, thereby performing a quantum gate between the two qubits \cite{reiserer_quantum_2014}. When a right-circularly polarized photon $\ket{\uparrow_z^p}$ is reflected off the cavity and the atom is in the state $\ket{\uparrow_z^a}$, strong coupling leads to a normal-mode splitting such that the photon is directly reflected from the first mirror. For all other state combinations, the photon enters the cavity and the combined atom-photon state thereby acquires a phase shift of $\pi$ \cite{duan_scalable_2004, reiserer_quantum_2014, Tiecke2014, Volz2014}. In the following, this quantum gate is used to experimentally implement a photon-to-atom state transfer as schematically depicted in Fig.\,\ref{fig:setup}. Toward this end, the atom is initially prepared in $\ket{\downarrow_x^a}$. The photonic state to be stored is $\ket{\varphi^p}=\alpha\ket{\downarrow_x^p} + \beta\ket{\uparrow_x^p}$. Upon reflection of the photon, the gate performs the following transformation:
$$\ket{\downarrow_x^a} \ket{\varphi^p} \rightarrow \frac{1}{\sqrt{2}} \left( \ket{\varphi^a} \ket{\downarrow_x^p} + iR_x^{\pi}\ket{\varphi^a} \ket{\uparrow_x^p} \right).$$
Here, $R_x^{\pi}$ denotes a rotation by $\pi$ around the x-axis and the atomic qubit is defined as $\ket{\varphi^a}=\alpha\ket{\downarrow_z^a}+\beta\ket{\uparrow_z^a}$. After the reflection process, the photonic polarization is detected in the x-basis, which unambiguously heralds successful photon-to-atom state transfer. However, compared to the input qubit, the resulting atomic state is rotated when the photon is detected in $\ket{\uparrow_x^p}$. In this case, quantum feedback in the form of a rotation $R_x^{\pi}$ is applied to the atom to deterministically recover the input qubit.

The experimental protocol is repeated at a rate of 1\,kHz. The atomic state is initialized via 140\,\textmu s of optical pumping into $\ket{\uparrow_z^a}$. Subsequently, the atom is rotated by $\pi/2$ into $\ket{\downarrow_x^a}$ using a pair of Raman laser beams, red-detuned by -0.15\,GHz from the D$_1$ line of $^{87}$Rb and applied for 1.7\,\textmu s. Then, a weak laser pulse with average photon number $\bar{n}=0.09$ and a Gaussian wave-packet envelope with a full width at half maximum of 0.6\,\textmu s is reflected from the cavity. The cavity is single sided (95\,ppm transmission of the coupling mirror; 8\,ppm combined scattering and absorption losses and transmission of the second mirror), such that $(70\pm2)\,\%$ of an incoming pulse resonant with the empty cavity are backreflected. On the relevant $\ket{2,2} \leftrightarrow \ket{3,3}$ transition of the D$_2$ line strong coupling is achieved (measured atom-cavity coupling constant $g=2\pi\times 6.7$\,MHz, atomic dipole decay rate $\gamma=2\pi\times 3.0$\,MHz, cavity field decay rate $\kappa=2\pi\times 2.5$\,MHz), with a reflectivity of $(66\pm2)\,\%$ on resonance. With the atom in $\ket{\downarrow_x^a}$ and the impinging photon in $\ket{\downarrow_x^p}$, this results in an average reflectivity of $(69\pm2)\,\%$. The conditional rotation of the atomic state by $\pi$ takes 3.4\,\textmu s. For analysis, the atomic state is rotated into one of three mutually unbiased bases and subsequently read out within 3\,\textmu s via cavity-enhanced fluorescence hyperfine state detection \cite{reiserer_arxiv_2014}.

\begin{figure*}
\includegraphics[width=2\columnwidth]{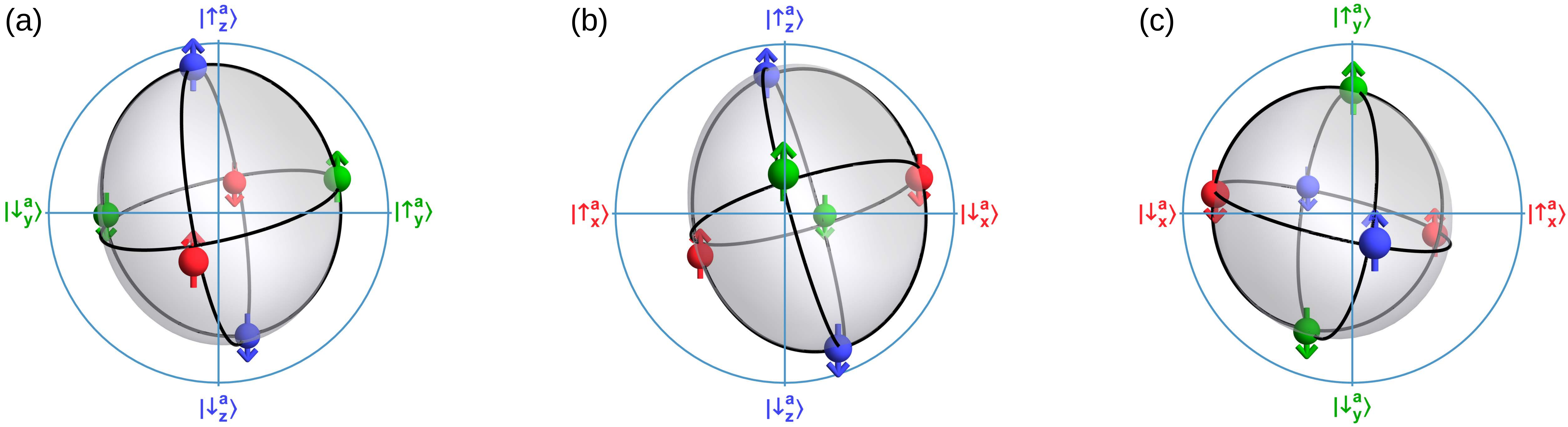}
\caption{\label{fig:storage}
(color online). Atomic quantum state after the storage process. (a) Front, (b) side, and (c) top view of the reconstructed Bloch sphere. Six photonic input states are mapped onto the atom, whose state is then reconstructed by using quantum state tomography (colored spheres with arrows). The kets on the unit circle indicate the ideal states. The full storage process is characterized via process tomography. The result is visualized by the gray sphere. The average state fidelity between the ideal and reconstructed atomic state is $\mathcal{F} = (86\pm1)\,\%$.}
\end{figure*}

To characterize the heralded storage, we perform quantum process tomography. To this end, we analyze the state transfer for six different input polarizations, forming three mutually unbiased bases. For each input polarization, we perform quantum state tomography on the atom. The result is depicted in Fig.\,\ref{fig:storage}. The average state fidelity is $\mathcal{F} = (86\pm1)\,\%$, with the individual fidelities $\mathcal{F}=\bra{\varphi^p}\rho_a\ket{\varphi^p}$ defined as the overlap between the density matrix of the measured atomic state $\rho_a$ and the ideal input qubit $\ket{\varphi^p}$. For heralding events where the photon was detected in $\ket{\downarrow_x^p}$ ($\ket{\uparrow_x^p}$), the average fidelity is $\mathcal{F}_{\ket{\downarrow_x^p}} = (84\pm1)\,\%$ ($\mathcal{F}_{\ket{\uparrow_x^p}} = (87\pm1)\,\%$), respectively. This value has to be compared to the best performance achievable with a classical device, which is 67.5\,\% \cite{specht_single-atom_2011} at the input photon number of $\bar{n}=0.09$ and the achieved efficiency of 39\,\%. The measured average state fidelity by far exceeds this limit, thereby proving the quantum nature of the device.

The role of the atom and photon in our scheme can be interchanged, which facilitates the transfer of a given atomic state onto the polarization of a photon. Because the atomic state detection is deterministic, the efficiency of this process is the product of the probability to have a single photon impinging onto the cavity and the reflectivity of the atom-cavity system. With a deterministic single-photon source, our system would therefore achieve 69\,\% efficiency. However, in the experiment, we instead employ weak laser pulses ($\bar{n}=0.08$).

\begin{figure}
\includegraphics[width=\columnwidth]{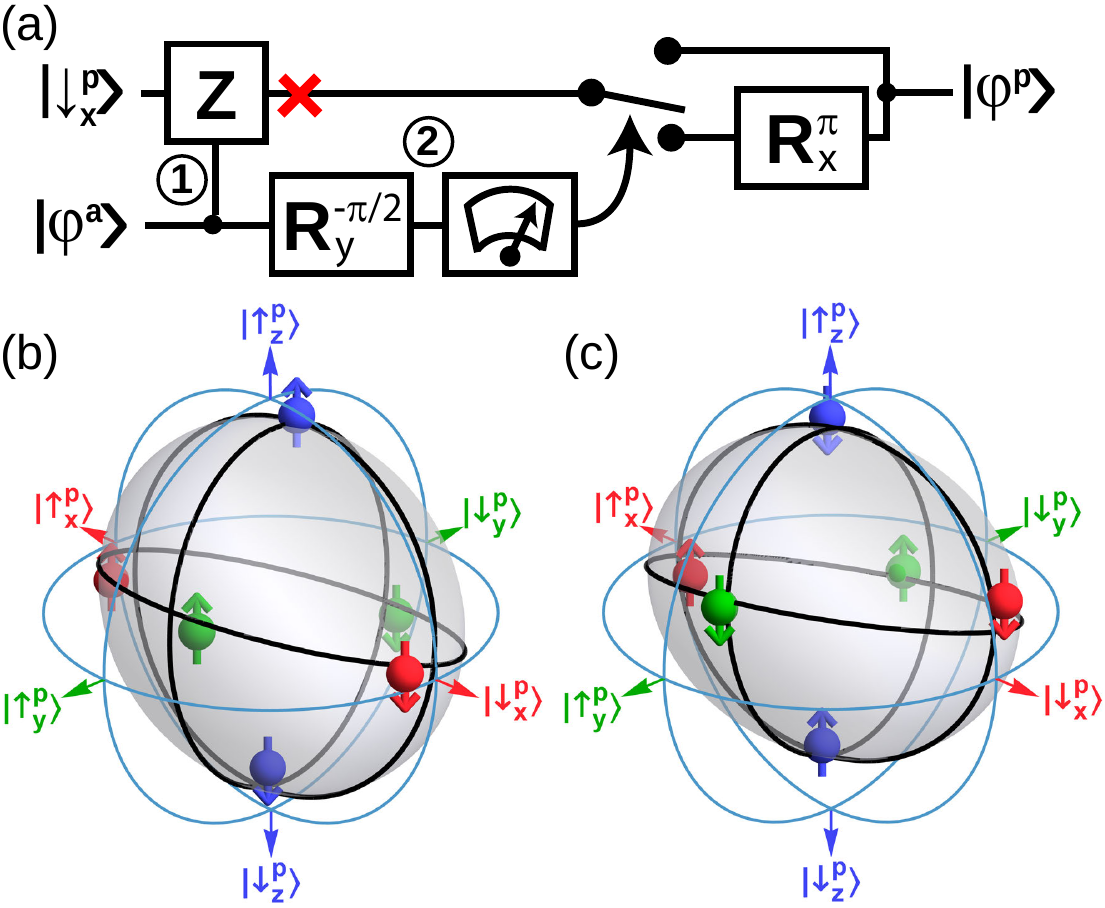}
\caption{\label{fig:readout}
(color online). Atom-to-photon state transfer: quantum circuit diagram and experimental results. (a) The atomic quantum state $\ket{\varphi^a}$ is to be mapped onto the polarization of an impinging photon. To this end, the photon is prepared in $\ket{\downarrow^p_x}$ and reflected from the atom-cavity system, thereby executing a controlled-phase quantum gate \textcircled{1}. Subsequently, the x-projection of the atomic state is measured by performing a $R_y^{-\pi/2}$ rotation followed by hyperfine state detection \textcircled{2}. When the measurement outcome is $\ket{\downarrow^a_x}$, the photon is projected into the desired state $\ket{\varphi^p}$, whereas the result $\ket{\uparrow^a_x}$ projects the photon to the rotated state $R_x^{\pi}\ket{\varphi^p}$. The red cross indicates that in our implementation the photon is detected before the measurement of the atomic state is performed, and the conditional polarization rotation is omitted.
(b) Results of the atom-to-photon state transfer conditioned on detection of the atom in $\ket{\downarrow^a_x}$. The gray reconstructed Poincar\'{e} sphere shows the process tomography results, with an average state fidelity $\mathcal{F}_{\ket{\downarrow^a_x}} = (88\pm1)\,\%$. (c) Photonic state when the atom is detected in $\ket{\uparrow^a_x}$. The rotation of the Poincar\'{e} sphere by $\pi$ around the x-axis is clearly visible. The average state fidelity with respect to the rotated state is $\mathcal{F}_{\ket{\uparrow^a_x}} = (85\pm1)\,\%$.}
\end{figure}

The protocol is depicted as a circuit diagram in Fig.\,\ref{fig:readout}(a). The photonic state is prepared in $\ket{\downarrow_x^p}$ and then reflected from the cavity. With the atom initially in the potentially unknown state $\ket{\varphi^a}=\alpha\ket{\downarrow_z^a}+\beta\ket{\uparrow_z^a}$, the reflection process results in
$$\ket{\downarrow_x^p} \ket{\varphi^a} \rightarrow \frac{1}{\sqrt{2}} \left( \ket{\varphi^p} \ket{\downarrow^a_x} + iR_x^{\pi} \ket{\varphi^p} \ket{\uparrow^a_x} \right).$$
Subsequent detection of the atomic state therefore projects the photon onto $\ket{\varphi^p}$ for a measurement result of $\ket{\downarrow^a_x}$, and onto $R_x^{\pi}\ket{\varphi^p}$ for $\ket{\uparrow_x^a}$. Postselection on $\ket{\downarrow^a_x}$, which occurs in 50\,\% of the cases where a photon has been reflected, completes the state transfer in our experiment. The result is shown in Fig.\,\ref{fig:readout}(b) in a Poincar\'{e}-sphere representation analogous to Fig.\,\ref{fig:storage}. We find an average state fidelity of $\mathcal{F}_{\ket{\downarrow^a_x}} = (88\pm1)\,\%$. When the atom is detected in $\ket{\uparrow^a_x}$, the photon ends up in a rotated state, as shown in Fig.\,\ref{fig:readout}(c). The average state fidelity with the expected rotated state $R_x^{\pi}\ket{\varphi^p}$ is $\mathcal{F}_{\ket{\uparrow^a_x}} = (85\pm1)\,\%$. Feedback onto the photonic polarization would again render the scheme deterministic. However, in the setting of a quantum network with heralded storage, direct rotation of the photonic polarization is not required to achieve a state transfer between remote atoms. Instead, it is fully sufficient to apply the conditional rotation only to the state which is stored in the memory node.

Several imperfections limit the fidelity of photon-to-atom and atom-to-photon state transfer in our experiment. The transverse mode of the incoming pulse has an overlap of $(92\pm3)\,\%$ with the cavity mode, reducing the average fidelity by 5.5\,\%. In the case of photon storage, further reduction arises from imperfect atomic state preparation (2.3\,\%), dark counts (1.2\,\%), finite quality of the polarization optics in the photonic state detection setup (0.7\,\%), two-photon events (1.2\,\%), and cavity birefringence (1\,\%). In total, these effects account for an average fidelity reduction of 12\,\%. Note, however, that the reduction depends on the photonic input state. This can be seen in the asymmetry of the reconstructed Bloch sphere in Fig.\,\ref{fig:storage}. In particular, storage of $\ket{\uparrow_z^p}$, which is mapped onto $\ket{\uparrow_x^a}$, has a lower fidelity than the other states, because the atom is initially prepared in the orthogonal state $\ket{\downarrow_x^a}$, in which it remains whenever the storage mechanism does not work as intended. An input-state-dependent analysis of the imperfections reproduces these findings.
In the case of atom-to-photon state mapping, the above-mentioned effects lead to an expected reduction of the average state fidelity of 11\,\% and 14\,\% for detection of the atom in $\ket{\downarrow^a_x}$ and $\ket{\uparrow^a_x}$, respectively. The former is smaller, because detection of the atom in $\ket{\downarrow^a_x}$ postselects on ideal atomic state preparation \cite{reiserer_quantum_2014}. By comparing the expected to the measured fidelities, we conclude that the mentioned effects include all dominant sources of error. None of the current imperfections is fundamental.

We emphasize that the presented storage scheme is largely insensitive to fluctuations of atom-cavity parameters, in particular to the atomic resonance frequency \cite{supplement} and the atom-cavity coupling strength \cite{duan_scalable_2004}. This relaxes the constraints on the trapping and cooling of the single atoms and potentially also on the indistinguishability of solid-state optical emitters, making our scheme promising for quantum state transfer and entanglement distribution between remote nodes in large-scale quantum networks. Even more important, the storage mechanism is robust with respect to the exact arrival time and fluctuations of the wave-packet envelope of the photon to be stored, as long as the photonic bandwidth is small compared to the cavity linewidth \cite{supplement, duan_scalable_2004}. This is in stark contrast to all previous approaches, especially state transfer based on coherent photon absorption \cite{specht_single-atom_2011, ritter_elementary_2012} and teleportation employing an optical BSM \cite{hammerer_quantum_2010, chen_memory-built-quantum_2008, bao_quantum_2012, nolleke_efficient_2013, gao_quantum_2013}.

This robustness of our scheme is particularly important for applications in long-distance quantum networks, where dispersion distorts the photonic wave packet and different path lengths, e.g., caused by changes in the ambient temperature that alter the length or refractive index of both fiber- and free-space channels, will lead to fluctuating photon arrival times. Paired with its versatility and unprecedented efficiency, our scheme thus brings the realization of global quantum networks using quantum repeaters \cite{briegel_quantum_1998} one step closer. Toward this end, the concept of quantum feedback, as introduced here to optical atom-cavity systems, could be combined with the transfer of quantum states and the entanglement of remote atoms. This could be implemented via the following two-step protocol: First, atom-to-photon state transfer or the generation of atom-photon entanglement is achieved, either via photon generation \cite{ritter_elementary_2012, reiserer_arxiv_2014} or by using the scheme demonstrated here; second, the heralded photon-to-atom state transfer at a remote network node will signal success of both protocol steps. The high rate of heralded remote entanglement which is achievable with this protocol will enable device-independent quantum key distribution \cite{Ekert2014} and loophole-free tests of quantum nonlocality \cite{Brunner2014}.

\begin{acknowledgments}
This work was supported by the European Union (Collaborative Project SIQS) and by the Bundesministerium f\"ur Bildung und Forschung via IKT 2020 (Q.com-Q) and by the DFG via NIM.
\end{acknowledgments}

\renewcommand{\thefigure}{S\arabic{figure}}
\setcounter{figure}{0}

\section*{Supplemental Material}
\begin{figure}[t]
\includegraphics[width=\columnwidth]{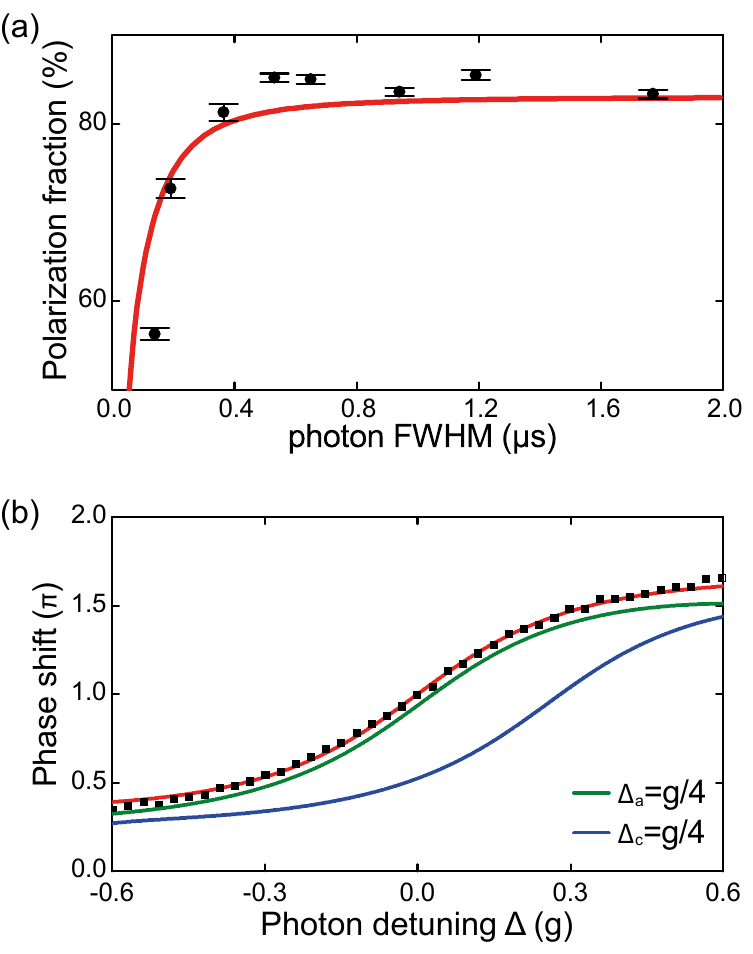}
\caption{\label{fig:phaseshift}
Robustness of the storage process with respect to relevant experimental parameters. (a) Bandwidth of the atom-photon interaction mechanism. Probability for observing an orthogonal linear polarization after reflection of photons with different width of the Gaussian temporal envelope. The mechanism works as intended when the polarization is flipped, i.e. for a photon duration above 0.5\,\textmu s. The red theory curve has the maximal success probability of the gate operation as the only free parameter. (b) Phase shift between left- and right-circular polarization with the atom in the coupled state $\ket{\uparrow_z^a}$. The detuning $\Delta=\omega_p-\omega_0$ of the photon frequency $\omega_p$ from the resonance frequency $\omega_0$ is given in units of the atom-cavity coupling strength $g$. The red solid line is the theoretical expectation deduced from an input-output treatment with atom (resonance frequency $\omega_a$) and cavity (resonance frequency $\omega_c$) on resonance ($\Delta_a=\omega_a-\omega_0=0, \Delta_c=\omega_c-\omega_0=0$). The theoretical curve for an atom (cavity) detuning of $g/4$ is shown as a green (blue) solid line.}
\end{figure}

In the following, the robustness of the presented storage mechanism with respect to fluctuating system parameters is demonstrated. We first investigate the influence of the bandwidth of the impinging photon. To this end, the atom is prepared in the coupled state $\ket{\uparrow_z^a}$, and faint laser pulses in the state $\ket{\downarrow_x^p}$ are reflected from the cavity. Their polarization is analyzed in the input (x-) basis. When the gate and thus the heralded storage mechanism works as intended, one expects to observe a polarization flip with unit probability. Figure \ref{fig:phaseshift}(a) shows the experimental results as a function of the full width at half maximum (FWHM) of the Gaussian temporal envelope of the impinging laser pulses. For pulses longer than about 0.5\,\textmu s, the probability is constant, which demonstrates that the storage mechanism is insensitive to the duration of the photon wave packet as long as this exceeds the cavity-field lifetime. The solid line is obtained using a theoretical model for the gate mechanism [S1] and calculating the expectation value of the phaseshift for a given FWHM of the photonic envelope. The only free parameter of this curve is the maximally possible flip probability, which is found to be 83\,\%.

More insight into the mechanism that determines the shape of this curve can be gained from Fig.\,\ref{fig:phaseshift}(b). Here, the phase shift for the coupled states (atom in $\ket{\uparrow_z^a}$, photon in $\ket{\uparrow_z^p}$) is depicted as a function of the frequency of the impinging photons. There is excellent agreement between the data (black squares) and the red solid line that has been calculated using input-output theory [S2, S3] and a steady-state Heisenberg-Langevin treatment [S1, S4] with the cavity offset frequency as the only free parameter. While the phase shift is $\pi$ exactly on resonance, it changes considerably over a detuning range of $0.1g$ corresponding to about one MHz in our experiment. This range is determined by the cavity linewidth $\kappa$ as long as $g \gg \kappa$ [S1]. Around zero detuning, the phase difference between the coupled and uncoupled-atom case can be approximated by a linear increase in the frequency domain, which corresponds to a constant shift of the photonic wave packet in the time domain. The gate mechanism starts to fail when this shift approaches the photon length, i.e. for shorter input pulses.

The storage mechanism is also only weakly dependent on the atom-cavity coupling strength $g$ and the atom-cavity detuning [S4]. The latter can be seen from the green theory curve in Fig.\,\ref{fig:phaseshift}(b), which was calculated for an atom detuned by a quarter of the atom-cavity coupling strength ($\Delta_a=g/4, \Delta_c=0$). The result hardly deviates from the curve on resonance (red). In contrast, shifting the cavity by the same amount ($\Delta_c=g/4$, $\Delta_a=0$) has more dramatic consequences (blue solid line). Consequently, the relative detuning between photon and cavity has to be kept well below the cavity field decay rate $\kappa$, something which is easily achievable with state-of-the-art cavity locking techniques.

\subsection*{References for Supplemental Material}
\begin{enumerate}[{[S1]}]
\setlength{\itemsep}{-\parsep}
\begin{small}
\item
A.~Reiserer and G.~Rempe, arXiv:1412.2889.
\item
C.~W. Gardiner and M.~J. Collett, Phys. Rev. A \textbf{31}, 3761 (1985).
\item
D.~F. Walls and G.~J. Milburn, \textit{Quantum Optics} (Springer, New York, 2008).
\item
L.-M. Duan and H.~J. Kimble, Phys. Rev. Lett. \textbf{92}, 127902 (2004).
\end{small}
\end{enumerate}
\clearpage

\end{document}